\pdfoutput=1
\documentclass[sigconf]{acmart}
\AtBeginDocument{%
  \providecommand\BibTeX{{%
    \normalfont B\kern-0.5em{\scshape i\kern-0.25em b}\kern-0.8em\TeX}}}

\setcopyright{acmcopyright}
\copyrightyear{2018}
\acmYear{2018}
\acmDOI{XXXXXXX.XXXXXXX}

\acmConference[Submitted to SIGCSE '24]{Submitted to 2024 ACM SIGCSE Technical Symposium}{August 20--23, 2023}{Portland, OR, USA}
\acmPrice{15.00}
\acmISBN{978-1-4503-XXXX-X/18/06}

\newcommand{\anonSemester}{Spring 2023}
\newcommand{\anonClassYear}{2023}

\newcommand{\anonCourseNum}{CS474}
\newcommand{\anonCourseTitle}{Intro Theoretical Computer Science}
\newcommand{\anonNumStudents}{41}

\newcommand{\probType}[1]{P#1}

\usepackage{colortbl}

\usepackage{tikz}
\usepackage{pgfplots}
\usepackage{pgfplotstable}

\usetikzlibrary{arrows.meta,automata,calc,fit,positioning,decorations.pathmorphing,decorations.text,decorations.pathreplacing,calligraphy,chains,shapes,matrix}
\tikzset{snake it/.style={decorate, decoration={snake,post length=0.5mm}}}
\tikzset{
	every state/.style={
		fill=gray!10,
		semithick
	},
	every edge/.style={
		draw,
		->,
		auto,
		very thick,
	},
	double distance=2pt,
	initial text={},
}

\pgfplotsset{compat=1.17}

\begin{document}

\title{Experiences with Research Processes in an Undergraduate Theory of Computing Course}


\author{Ryan E. Dougherty}
\email{ryan.dougherty@westpoint.edu}
\orcid{0000-0003-1739-1127}
\affiliation{%
  \institution{United States Military Academy}
  \city{West Point}
  \state{New York}
  \country{USA}
  \postcode{10996}
}

\renewcommand{\shortauthors}{R. E. Dougherty}

\begin{abstract}
Theory of computing (ToC) courses are a staple in many undergraduate CS curricula as they lay the foundation of why CS is important to students.
Although not a stated goal, an inevitable outcome of the course is enhancing the students' technical reading and writing abilities as it often contains formal reasoning and proof writing. 
Separately, many undergraduate students are interested in performing research, but often lack these abilities.
Based on this observation, we emulated a common research environment within our ToC course by creating a mock conference assignment, where students (in groups) both wrote a technical paper solving an assigned problem and (individually) anonymously refereed other groups' papers.  
In this paper we discuss the details of this assignment and our experiences, and conclude with reflections and future work about similar courses. 
\end{abstract}



\received{20 February 2007}
\received[revised]{12 March 2009}
\received[accepted]{5 June 2009}

\maketitle

\section{Introduction}

Many CS undergraduate curricula contain at least one course on theory of computing (ToC) topics, whose goal is to show students how universally applicable a set of core, formal concepts is to all of their previous studies. 
This is often the most challenging course in the curriculum, in part due to the high amount of formal writing (proofs and otherwise) and reasoning involved. 
Often the only course that is a prerequisite to ToC is discrete mathematics, which has the job of introducing proof writing and mathematical reasoning. 
At many institutions (including ours), however, there is a large temporal separation between when most students take discrete math and ToC. 
Additionally, students may be able to write and understand proofs of some claim well, but struggle to put that proof within an appropriate context and derive conclusions based on their proof.
Further, they may be able to read their proofs well (as they wrote them), but may have trouble reading another student's proof. 

Therefore, in \anonSemester, we created a semester-long assignment to have students undergo a realistic scenario that researchers regularly experience that emphasizes writing and reading.
Specifically, we created a double-blind ``mock conference'' in which students were assigned a problem in groups, created an anonymously presented technical paper, and (anonymously) individually refereed other student submissions. 
We had two goals for this assignment:
\begin{itemize}
    \item determine how well students can write technical papers within an undergraduate ToC course on topics within that course; and 
    \item determine how well students give feedback to other student paper submissions in terms of accuracy and completeness.
\end{itemize}

In this paper we will give our experiences in designing such an assignment, what happened during our course offering with this assignment, and our recommendations for educators in implementing a similar assignment.
Our study has been approved by our institutional review board.
The rest of the paper is as follows.
In Section~\ref{sec:related_work} we list related work.
In Sections~\ref{sec:course_context} and \ref{sec:assignment_design} we give the design of the assignment.
Section~\ref{sec:issues_and_challenges} contains the issues and challenges we faced during and after the assignment.
Section~\ref{sec:reflections} contains our reflections, including what worked for us and what did not.
We enumerate future work and recommendations in Section~\ref{sec:future_work}.
Finally, we conclude in Section~\ref{sec:conclusion}.

\section{Related Work}\label{sec:related_work}

There have been several published works on simulating a research-like environment in undergraduate courses; namely, in publishing an anonymous paper to a ``mock'' conference, and refereeing submissions. 
Tanner \cite{tanner2015learning} had such an environment in a biomedical sciences course. 
Budny et al. \cite{budny2002simulated} created a mock professional conference for engineering freshmen. 
The setup of Keesee and Bauer-Reich \cite{keesee2022using} was similar to ours but with an engineering course instead. 
However, their course was more focused on the conference itself (as a part of its lectures), rather than its being an out-of-class assignment.
Hadfield and Schweitzer \cite{hadfield2009building} integrate research experiences throughout their CS undergraduate curriculum.
Mallet \cite{mallet2005authentic} created such a conference for an undergraduate mathematical modelling course. 
Kumar \cite{kumar20114} creates a mock conference model for teaching undergraduates in general. 
Cass and Fernandes \cite{cass2008simulated} presents a model similar to ours except that students are told that their papers will be reviewed by an ``outside'' committee. 

None of the above works include the results of refereeing submissions from other students. 
Jones et al. \cite{jones2006undergraduate} had undergraduate students referee submissions from a ``real'' undergraduate journal in neuroscience, but does not include submitting works (whether a ``real'' or ``mock'' conference).

\section{Course Context}\label{sec:course_context}

In \anonSemester\ we had three sections of our ToC course, titled \anonCourseTitle, all taught by the author, with a total of \anonNumStudents\ students.
Our ToC course is primarily taken during the students' junior year, with prerequisites being discrete mathematics and digital logic, and a corequisite of algorithms; ToC is only offered once per year. 
Importantly, students for this particular offering had some training in \LaTeX\ during their discrete math course, specifically within Overleaf.
There are no formal post-requisites, but students very often take a year-long capstone course and an Operating Systems course in the next semester; see Section~\ref{sec:future_work} for details.

Most ToC courses are divided into four large sections: (1) regular languages, (2) context-free languages, (3) Turing Machines and decidability, and ending with (4) undecidability. 
The order of the vast majority of such offerings is $(1) \to (2) \to (3) \to (4)$, and our course follows the same roadmap other than including NP-completeness after (4).
The first two sections contain a formal model to both define and analyze, and some amount of proofs to show that each model cannot solve ``all'' problems; the fourth solely contains proofs that Turing Machines--computationally equivalent to ``real'' computers--cannot solve every problem, or even ``reasonable'' problems.
There are two goals of dividing the course into sections like this: build upon previous knowledge in terms of the model's formal definition and its capabilities, and to showcase the classic trade-off of increased computational power vs. tractability of asking questions about the model. 
For example, it is algorithmically undecidable to determine if two machines in (2) have the same behavior, whereas it is possible for two machines in (1).

Student assessment for our course was divided up as follows:
\begin{itemize}
    \item 5 In-Class Group Presentations: 200 points total
    \item 10 In-Class Quizzes: 150 points total
    \item Paper Writing (Part 1 of conference): 75 points
    \item Paper Refereeing (Part 2 of conference): 75 points
    \item Formal Group Presentation (NP-Completeness): 150 points
    \item Final Exam: 250 points
    \item Lesson Preparedness: 100 points
\end{itemize}
These choices were to support the conference with a nonnegligible amount of the final grade while also not dissuading students from not being prepared for lessons, studying for the final exam, or preparing for the formal presentations.

\section{Assignment Design}\label{sec:assignment_design}

This section contains our process in designing the paper writing and refereeing assignments.
Students were put into groups of two or three (potentially spanning across the three sections) randomly by the instructor.
The assignment naturally is broken into three parts: paper writing (Part 1), paper refereeing (Part 2), and the conference itself (Part 3).
The first two parts were conducted entirely through EasyChair.\footnote{\url{https://easychair.org/}}

\subsection{Paper Writing (Part 1)}
For Part 1, groups were required to use the ACM SIG Proceedings \LaTeX\ template for standardization purposes; a link to the Overleaf template was provided for ease of student use. 

Each group is assigned a unique problem that was manually created by the instructor that has some real-world application in mind.
For example, one problem was to create a context-free grammar for edit distance of strings, as that problem has applications in computational biology.  
Each problem potentially had several sub-problems; the problem categories and what was required for each are given next.
\begin{itemize}
    \item[\probType{1}] -- construction of some formal object, along with a formal definition of the object and proof that the object's language is correct.
    An example problem could be ``Construct a non-deterministic finite automaton for the language $L = \{...\}$'' or ``Show that $L = \{...\}$ is context-free by creating a context-free grammar for it.''
    \item[\probType{2}] -- showing that a language operation is correct. 
    Students need to consider an arbitrary object with that language, how to change that object for the operation, and to show that the resulting object has the desired language. 
    An example problem could be ``Let $L$ be a regular language, and let $s(L)$ be the set of strings in $L$ with last character removed (if there is one).
    Show that if $L$ is regular, then $s(L)$ is also regular.''
    \item[\probType{3}] -- showing that a language is not regular using the Pumping Lemma (PL) for Regular Languages. 
    We provided students the steps for any proof using the PL during previous lessons, such as picking an arbitrary string in the language, performing all decompositions of the string according to the PL's rules, and finding a repetition that leaves the language.
\end{itemize}

Table~\ref{tbl:num_assigned_per_category} shows the number of groups assigned for each problem category.
Note that some groups have sub-problems that span several categories. 
The instructor made an effort to equally distribute paper categories across groups. 

The grade breakdown for Part 1 is as follows, with a total of 75 points (plus 10 bonus points for accepted papers):
\begin{itemize}
    \item[2 pts] Paper is anonymized; if not done, the paper is not graded.
    \item[3 pts] Appropriate title.
    \item[5 pts] Appropriate abstract (both length and content).
    \item[5 pts] Inclusion of relevant background material.
    \item[5 pts] Inclusion of any necessary notation and definitions.
    \item[30 pts] Accuracy of proof/argument/construction.
    \item[10 pts] Appropriate discussion/future work/conclusion sections.
    \item[10 pts] Professional writing style and formatting.
    \item[5 pts] References are properly formatted.
\end{itemize}

\begin{table}
\caption{\label{tbl:num_assigned_per_category}Number of groups assigned per problem category for Part 1.}
\begin{tabular}{|l|l|l|l|}
\hline
Problem Type & \probType{1} & \probType{2} & \probType{3} \\ \hline
Number of Groups    & 10 & 9 & 6 \\ \hline
\end{tabular}
\end{table}

As nearly all students have not performed research at this point, instructions were provided on how to appropriately structure a research paper: how to write an abstract (at most 200 words); introduction section; background and related work section that includes applications and motivation; a section for any necessary formal definitions; theorems within provided proof environments; and sections for discussion, future work, and concluding the paper.
For background and related work, students were advised to search Google Scholar with appropriate keywords and read abstracts of ``similar'' papers; if any are sufficiently similar, then they were to read any relevant theorems and proofs within it for inclusion and comparison in their own paper. 
For the discussion section, students were not required but encouraged to think more deeply about the implications of their results, and if they can be extended or generalized. 

A rubric was provided that assigned many of the points to correctness of proofs, but a nontrivial amount for the other sections, anonymization, and properly formatted references. 
Additionally, an incentive of bonus points was provided for papers that get accepted to the conference.
Groups were given approximately a month and a half to complete the paper, and was assigned roughly three weeks into the course. 
The problems had a difficulty that was appropriate for a regular (individual) homework spanning two or three weeks, and thus students were expected to easily complete solving the problem within the allotted time.
Although there was no scaffolding to Part 1, the instructor provided a recommended schedule for when each part of the paper should be done. 

\subsection{Paper Refereeing (Part 2)}
For Part 2 students worked individually for (anonymously) refereeing other papers from the conference.
Instructions were provided for how to write an appropriate review for a paper, as all students have not reviewed for a conference before.
All students were assigned two papers to review (assigned randomly by the instructor), and most of the points are for accuracy of the evaluation.
The overall evaluation (a numerical score) students assigned to a paper was graded on how similar it was to the instructor's score of the same paper.
The instructor told students that the accuracy and honesty of their reviews are what primarily counted for their grade.
Because most students have not read a technical paper before, the instructor give a link to the paper of Keshav \cite[Section 2]{keshav2007read}.
Students had approximately three weeks to complete this part.

While students were refereeing, the instructor graded each of the submitted papers, and made a decision for each as to whether it should be accepted into the conference. 
At the end of the review process, the instructor's grade and feedback, along with the anonymous student reviews, were provided to each of the groups.
The grade breakdown for Part 2 is as follows; the same rubric is used for both reviewed papers, for a total of 75 points.
\begin{itemize}
    \item[5 pts] Overall evaluation: a score from -3 (strong reject) to +3 (strong accept).
    \item[2.5 pts] Brief paper summary.
    \item[2.5 pts] Evaluation summary.
    \item[25 pts] Evaluation.
    \item[2.5 pts] Reviewer's confidence: a score from 1 to 5 about how well the reviewer understood the paper. 
\end{itemize}

\subsection{Conference (Part 3)}

For groups of accepted papers, the instructor notified them that they need to incorporate changes given by the instructor and reviewers within two weeks (along with de-anonymizing their paper).
Once the accepted paper revisions were taken, the instructor assembled them into one PDF to share with the rest of the course (with permission of the students in these groups).
The conference was held in the last class period of the semester, during which students of accepted papers shared their story about how they approached the problem, what worked and what did not, and their recommendations for the rest of the students in tackling similar problems. 

\section{Issues and Challenges}\label{sec:issues_and_challenges}

This section contains the issues and challenges we faced developing and running our mock conference.
Since this project took the large majority of the semester, any assigned projects had to only contain course concepts from the first few sections of the course, namely regular languages (1) and context-free languages (2). 
The section on undecidability (4) is entirely proof-based, and thus the assigned problems have a bias in that there are fewer proofs than if a different ordering on course concepts were imposed.

Additionally, it was the instructor's burden to design problems that are not only unique to groups, but are not readily available online or elsewhere; these constraints were to avoid potential plagiarism.

In Part 1, there was no scaffolding. 
As a result, three groups did not submit their paper on time, and thus were not eligible to be accepted to the conference nor to be reviewed by other students; their work was nonetheless graded with the same rubric and feedback by the instructor. 

\section{Reflections}\label{sec:reflections}

This section contains our reflections after running the conference.
We did not create any student surveys, but did store the following data: individual student IDs (made anonymous by a faculty member not part of the analysis), a mapping between student IDs and group identifiers, scores for each group for each of the items in Section~\ref{sec:assignment_design} (both Parts 1 and 2), conference acceptance scores, and all text for reviews in Part 2.  
All data has been anonymized by the same faculty member.

\subsection{What Worked Well}
Having real-world applications as a requirement for generating the problems helped students see the purpose of such problems.
Additionally, since students had to perform background research, they were largely able to contextualize what is fundamentally different about their problem compared to previous work.

The tools used made coordination between the instructor and groups very simple and efficient.
Having a \LaTeX\ template along with a platform for distributing and compiling the document repository was instrumental in making this assignment not only possible, but realistic enough to feel like a ``real'' conference. 

Taking the time to generate different categories of problems based on real-world applications made the conference proceedings more realistic, as that emulates ``real'' conference proceedings with a wide body of research topics. 
Anecdotally during the conference (Part 3), we found that students were curious about how other groups approached their problems, as they were often quite different from their own.

Part 2 of the conference showed that students were in general honest about their peers and themselves.
However, some lower-performing students often would praise a paper and then give a low confidence score in their abilities to understand the paper (or vice versa).

\subsection{What Did Not Work Well}

One of the issues with EasyChair's free version is that it is limited in how many submissions are allowed. 
Even if a group accidentally made a submission that they want removed (e.g., forgetting a group member), or the instructor makes a test submission and subsequently removes it, these count against the total number of submissions anyways.
We did not seek funding from our department for using a paid version, nor looked for a free alternative.
We return to this as a recommendation in Section~\ref{sec:future_work}.

Part 1's length was a major issue, as students are often tempted to not start an assignment until close to its due date. 
As performing nearly any kind of research takes more time than predicted, nearly all groups made their first submission (before any modifications) within a day of the due date.
EasyChair by-default allows groups to submit their work as many times as they want before the deadline, but most groups did not take advantage of this. 

Students are known to only care about assignments if they are worth a significant number of points \cite{lang2021small}. 
The total number of points for the two parts was 150, which is 15\% of the total student grade.
Even though this was a semester-long project, the fact that it was worth only this many points most likely negatively influenced how much dedication students had to writing the paper. 

Even though the instructor encouraged students not to spend significant time on the paper's visual aesthetics (and to instead save this once everything else is done), students ended up not following this advice. 
All groups with a problem of type \probType{1} (creating a formal object) used some kind of visualization tool for showing the formal object to the reader, and then later proving its correctness.
Students were allowed to use any tool that they wanted, either hand-drawn or on a computer. 
One tool the instructor suggested (among several) was {\tt tikz-automata}, which involves \LaTeX\ code to produce automata figures, as was done for the course's lecture notes; see Figures~\ref{fig:tikz_example} and \ref{fig:tikz_example_code} for an example and provided sample code.\footnote{We provided students a \LaTeX\ header that contains defined commands for {\tt automatonCenter} to maintain consistency.}
Many groups opted to use this tool, but ended up spending a significantly more than expected amount of time learning how to use it. 

\begin{figure}
    \centering
    \caption{\label{fig:tikz_example}A {\tt tikz-automata} example within our ToC course.}

    \begin{tikzpicture}[node distance=2cm]
    \node[state,initial] (1)   {$q_0$}; 
    \node[state,accepting] (2) [right=of 1] {$q_1$}; 
    \path[->] 
    (1) edge [bend left] node {$0$} (2)
    edge [loop above] node {$1$} ()
    (2) edge [bend left] node {$1$} (1)
    edge [loop above] node {$0$} ();
\end{tikzpicture}
\end{figure}
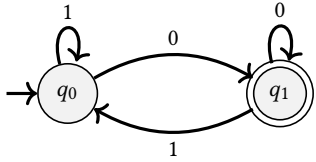

\begin{figure}
    \centering
        \caption{\label{fig:tikz_example_code}Code for the automaton in Figure~\ref{fig:tikz_example}.}
\begin{verbatim}
\automatonCenter{
    \node[state,initial] (1)   {$q_0$}; 
    \node[state,accepting] (2) [right=of 1] {$q_1$}; 
    \path[->] 
    (1) edge [bend left] node {$0$} (2)
    edge [loop above] node {$1$} ()
    (2) edge [bend left] node {$1$} (1)
    edge [loop above] node {$0$} ();
}
\end{verbatim}

\end{figure}

The paper acceptance criteria of other reviewers did not necessarily match those of the instructor.
The reviewing process used a -3 (strong reject) to +3 (strong accept) scale for papers; this is the default on EasyChair. 
Group paper scores had an average of slightly above 1.0.
On average, there was a 1.65 point difference between the average student score and that of the instructor for each submitted paper. 
Additionally, the instructor gave the anonymized papers to a colleague\footnote{This colleague has not ever taught ToC, but is aware of some of the concepts and proofs.} for determining acceptance; there was a 1.55 point difference there (including the three papers submitted late). 
Table~\ref{tbl:avg_rating_comparison} contains the rating results for the ten accepted papers from the instructor, the student average rating for that paper, and that of the colleague.
The difference between the highest and lowest student ratings are also given. 
Note that many of the papers have a very large student delta, especially group \#17 with a delta of 5. 
This was primarily due to students' reviews containing praise for the quality of writing and not about proof/argument accuracy.

Understandably, some students did not follow some of the instructor's recommendations. 
For \LaTeX, many groups did not use theorem and figure environments to structure their proofs, even though the instructor provided examples of how to use them.
In several cases, groups would directly copy and paste some lecture notes for the definition and notation section (with appropriate citation).

\begin{table*}
\caption{\label{tbl:avg_rating_comparison}Instructor Rating, student average rating, and colleague rating for the 10 accepted papers to the conference (given by group ID).
Groups are sorted left-to-right in decreasing order of total score given by the instructor.
Colors are given as follows: blue indicates 3.0, green is 2.0 to 2.9, yellow is 1.0 to 1.9, orange is 0.0 to 0.9, dark orange is -1.0 to -0.1, red is -2.0 to -1.9, and gray is -3.0 to -2.1.
Additionally, the difference between the highest and lowest student ratings is also given. 
}
\begin{tabular}{|l|l|l|l|l|l|l|l|l|l|l|}
\hline
Group ID               & 3                           & 7                           & 8                           & 15                          & 10                          & 11                          & 4                           & 12                          & 17                          & 16                          \\ \hline
Instructor Rating      & \cellcolor[HTML]{68CBD0}3   & \cellcolor[HTML]{68CBD0}3   & \cellcolor[HTML]{68CBD0}3   & \cellcolor[HTML]{F8FF00}1   & \cellcolor[HTML]{34FF34}2   & \cellcolor[HTML]{34FF34}2   & \cellcolor[HTML]{34FF34}2   & \cellcolor[HTML]{34FF34}2   & \cellcolor[HTML]{F8FF00}1   & \cellcolor[HTML]{F8FF00}1   \\ \hline
Student Average Rating & \cellcolor[HTML]{34FF34}2.5 & \cellcolor[HTML]{F8FF00}1.2 & \cellcolor[HTML]{FFCC67}0.0 & \cellcolor[HTML]{34FF34}2.0 & \cellcolor[HTML]{FFCC67}0.0 & \cellcolor[HTML]{F8FF00}1.2 & \cellcolor[HTML]{F8FF00}1.4 & \cellcolor[HTML]{FFCC67}0.2 & \cellcolor[HTML]{F8FF00}1.4 & \cellcolor[HTML]{FFCC67}0.5 \\ \hline
Student Max - Min      & 1                           & 4                           & 3                           & 2                           & 3                           & 1                           & 4                           & 3                           & 5                           & 3                           \\ \hline
Colleague Rating       & \cellcolor[HTML]{F8FF00}1   & \cellcolor[HTML]{34FF34}2   & \cellcolor[HTML]{F8A102}-1  & \cellcolor[HTML]{34FF34}2   & \cellcolor[HTML]{FFCC67}0   & \cellcolor[HTML]{FE0000}-2  & \cellcolor[HTML]{34FF34}2   & \cellcolor[HTML]{F8A102}-1  & \cellcolor[HTML]{FFCC67}0   & \cellcolor[HTML]{34FF34}2   \\ \hline
\end{tabular}
\end{table*}

\subsection{Limitations}

It is difficult to attribute an instructor grade on the accuracy of an argument to all of the merits of a student paper.
Published research papers, even within technical venues, are often worth more than just their technical arguments. 
Especially with student writing, the quality of exposition, figures, flow, and adequate background research are also important to the success of a paper. 
Therefore it is possible that a paper was not accepted to the conference that should have been (or vice versa) based on the grade breakdown the instructor chose. 

Certain biases of the instructor, if they existed, could have altered both of the parts, within the instructions and grading.
For example, the increased stressing of proof accuracy by the instructor could have led some groups to overemphasize that section, or potentially having lesser performing groups ignoring the proof section altogether (so that they can obtain points that they can more reasonably get).

It is nearly impossible to give an assignment like this in an earlier semester (at least at our institution), nor within a course that is not technical in nature. 
We believe it is possible to modify Part 1's writing component to be more expository or survey-based and include this assignment in, for example, a CS ethics course. 

Due to lack of time, Part 3 (i.e., the conference itself) during the last class period did not include full-length formal presentations.
We did not feel this to be necessary as all students had five formal presentations throughout the semester already.
However, having a true research-style talk would be the final part of a standard conference and research experience.

\section{Future Work \& Recommendations}\label{sec:future_work}

This section contains future work already in progress, and recommendations for practitioners. 

The CS major stresses not only technical reading and writing, but also communication to audiences, both technical and not.
To this end, we created another set of assignments in the same course and semester in which students gave five oral presentations on course concepts throughout the semester. 
Another motivation for having technical presentations is a follow-on capstone course, which takes place across two semesters.
In that capstone course, students are placed into larger groups (between 4 and 6 students each) and are assigned a computing problem to solve for a client. 
Throughout the two semesters, students must make several technical presentations; for juniors who took our ToC course, we will perform a statistical analysis between how well those students give oral presentations and those who did not take our course with our paper assignment. 
Another senior-level course at our institution is Operating Systems, and we are currently designing a paper writing assignment based on performance analysis of writing a OS component in {\tt C}.

Based on the previous paragraph and the work in this paper, we created research questions that we are in the process of addressing in future work:
\begin{itemize}
    \item RQ1: Does writing a research-style paper on a topic in an undergraduate ToC course (i.e., Part 1) improve performance on that topic (e.g., exam scores) more than for other topics? Additionally, if they scored a B or better in Part 1, is there a stronger correlation?
    \item RQ2: Same as RQ1, but for reviewing a paper.
    \item RQ3: Same as RQ2, but for reviewing a paper in the same paper category as that of their paper in Part 1.
    \item RQ4: Do undergraduate students reviewing fellow students' papers review primarily on merit or readability?
    \item RQ5: How do undergraduate students, who have participated in our paper writing assignment, perform in follow-up courses that have technical presentations or paper writing assignments, compared to students who did not?
    \item RQ6: Same as RQ5, but instead comparing presentations within our ToC course and those in the follow-up courses.
\end{itemize}

\begin{figure}
\caption{\label{fig:accuracy_vs_review_quality} Argument Accuracy for Part 1 vs. Review Evaluation (total) for Part 2 (as percentages).}
\centering
\begin{tikzpicture}
\begin{axis}[%
xlabel={Argument Accuracy (Part 1)},
ylabel={Review Evaluation (Part 2)},
legend pos=north west,
scatter/classes={%
    a={mark=o,draw=rred},
    b={mark=x,draw=bblue}}]
\addplot[scatter,only marks,%
    scatter src=explicit symbolic]%
table {
x y	
33.33333333	84
61.66666667	76
91.66666667	86
33.33333333	80
82.5	94
33.33333333	86
88.53333333	84
49.16666667	82
63.33333333	84
56.66666667	76
93.33333333	90
74.16666667	86
49.16666667	88
88.53333333	88
86.66666667	100
89.06666667	80
46.66666667	28
90.16666667	80
82.5	72
91.66666667	86
90.16666667	82
98.33333333	82
61.66666667	98
46.66666667	88
89.06666667	98
58.86666667	78
75	90
75	82
81.66666667	74
86.66666667	96
63.33333333	84
56.66666667	88
62.66666667	78
79.16666667	94
62.66666667	86
58.86666667	82
98.33333333	92
74.16666667	70
93.33333333	80
79.16666667	98
81.66666667	92
};
\end{axis}
\end{tikzpicture}
\end{figure}
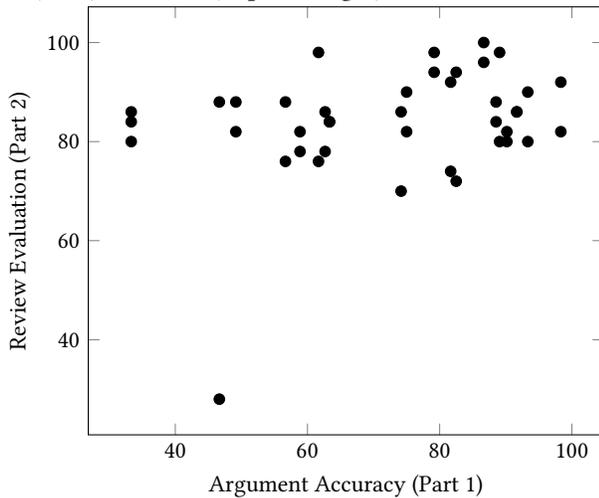

In spirit of partially addressing some of these questions (particularly RQ2 and RQ3), observe Figure~\ref{fig:part1_vs_part2}, a scatter plot of Part 1 vs. Part 2 overall scores; and Figure~\ref{fig:accuracy_vs_review_quality}, a scatter plot of only the correctness of the paper's main argument and the review evaluation quality.
Apart from one student, there does not appear to be any prediction between how well students can write a technical paper vs. how well they can provide a review of one; nor is there a correlation between the ``main'' components of both parts of the assignment.
However, more statistical analysis of the grades and empirical analysis of the text in paper/review submissions need to be performed.

\begin{figure}
\caption{\label{fig:part1_vs_part2} Overall Part 1 vs. Part 2 Scores (as percentages).}
\centering
\begin{tikzpicture}
\begin{axis}[%
xlabel={Part 1 Score},
ylabel={Part 2 Score},
legend pos=north west,
scatter/classes={%
    a={mark=o,draw=rred},
    b={mark=x,draw=bblue}}]
\addplot[scatter,only marks,%
    scatter src=explicit symbolic]%
table {
x y	
39.6	86.66666667
62.7	77.66666667
86.66666667	85
39.6	80
83.33333333	90.66666667
39.6	85.33333333
88.33333333	82
71.33333333	80.66666667
75.33333333	83.66666667
76.33333333	81.66666667
91	85.33333333
79.33333333	84.33333333
71.33333333	84.66666667
88.33333333	89.33333333
86.33333333	99.33333333
91.96	82
58.8	44
88.73333333	78.33333333
83.33333333	69.66666667
86.66666667	86.66666667
88.73333333	83.33333333
95	81.33333333
62.7	96.66666667
58.8	87.66666667
91.96	95
62.88	78.66666667
85	85.33333333
85	86.33333333
77	75.66666667
86.33333333	91
75.33333333	83.33333333
76.33333333	84.66666667
69.73333333	77.66666667
87.33333333	94.33333333
69.73333333	87.33333333
62.88	79.66666667
95	91.66666667
79.33333333	74
91	80.66666667
87.33333333	97
77	92.33333333
    };
\end{axis}
\end{tikzpicture}
\end{figure}
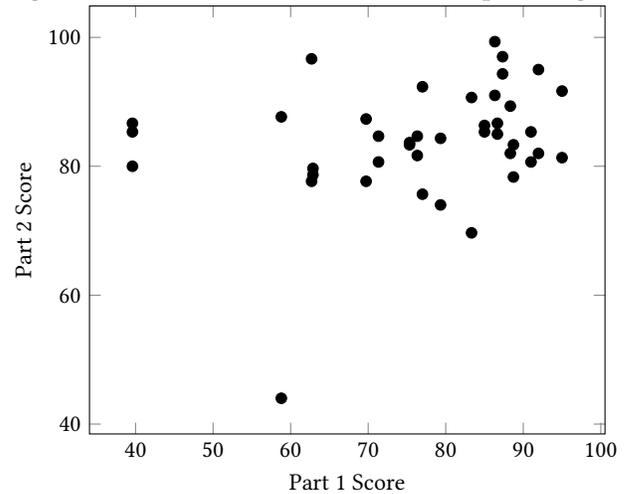

For practitioners looking to implement a similar assignment, we recommend using a publically available \LaTeX\ guide as it carries a steep learning curve.
If the course is not entirely technical, then we do not see any reason to use \LaTeX\ within an undergraduate classroom setting.
Additionally, we recommend practitioners to scaffold Part 1, as that in practice takes much more time and effort to complete than Part 2.

We recommend using a free alternative to EasyChair, especially for larger courses than ours, as its limitations for the free version influenced how we designed our assignment.
One possibility is HotCRP\footnote{\url{https://hotcrp.com/}}, which is used by many ``real'' CS conferences. 

Finally, several students have performed research before the start of this course (including one who has presented at a ``real'' CS conference), and thus impose a bias on the data: their writing is likely higher quality, and so their papers are more likely to be accepted.
We propose extending some or all of the research questions above to compare students who have performed research previously vs. those who did not.
The effects such students have we believe would be small as the number of undergraduate students who have presented research by the middle of their junior year is generally small.

\section{Conclusion}\label{sec:conclusion}

In this paper we gave our experience in creating and running an assignment within a ToC course that emulates a CS conference.
Students were grouped to solve a unique assigned problem and asked to write a technical paper in \LaTeX.
Further, they individually read and anonymously provided feedback to some other group submissions. 
We discussed some of the issues and challenges in creating such an assignment.
We hope that this paper gives CS educators inspiration to develop a similar assignment for other CS courses, and to determine the effectiveness of having research-style assignments within them.

\begin{acks}
We would like to thank the \anonCourseNum\ students in \anonClassYear\ for participating in our study and supporting CS research.
We also thank Dr. Maria Ebling for guidance in designing the future work research questions and studies, and aiding in the IRB process. 
The opinions in the work are solely of the author, and do not necessarily reflect those of the U.S. Army, U.S. Army Research Labs, the U.S. Military Academy, or the Department of Defense.
\end{acks}

\bibliographystyle{ACM-Reference-Format}
\bibliography{sample-base}

\end{document}